\documentclass{article}
\usepackage{spconf,amsmath,graphicx}
\usepackage{echodefs}

\newcommand{\rev}[1]{{\color{black}#1}}

\usepackage{import}
\usepackage{transparent}
\usepackage{booktabs}
\usepackage{multirow}
\usepackage{bm}
\usepackage[normalem]{ulem}

\newcommand{\sfr}{\mathsf{r}}
\newcommand{\sfy}{\mathsf{y}}

\title{Fast Optical System Identification by Numerical Interferometry}
\name{Sidharth Gupta$^*$, R\'emi Gribonval$^\dagger$,
Laurent Daudet$^\ddagger$ and Ivan Dokmani\'c$^*$\thanks{$^*$ S. Gupta and I. Dokmani\'c would like to acknowledge the support from the National Science Foundation under Grant CIF-1817577}}
\address{$^*$University of Illinois at Urbana-Champaign; $^\dagger$Univ Lyon, Inria, CNRS, ENS de Lyon; $^\ddagger$LightOn, Paris \\
{\small \tt gupta67@illinois.edu, remi.gribonval@inria.fr, laurent@lighton.ai, dokmanic@illinois.edu}}
\begin{document}
\ninept
\maketitle
\begin{abstract}

We propose a numerical interferometry method for identification of optical multiply-scattering systems when only intensity can be measured.  Our method simplifies the calibration of optical transmission matrices from a quadratic to a linear inverse problem by first recovering the phase of the measurements. We show that by carefully designing the probing signals, measurement phase retrieval amounts to a distance geometry problem---a multilateration---in the complex plane. Since multilateration can be formulated as a small linear system which is the same for entire rows of the transmission matrix, the phases can be retrieved very efficiently. To speed up the subsequent estimation of \rev{transmission matrices}, we design calibration signals so as to take advantage of the fast Fourier transform, achieving a \rev{numerical} complexity almost linear in the number of \rev{transmission matrix entries}.
We run experiments on real \rev{optical} hardware and use the \rev{numerically} computed transmission matrix to recover an unseen image behind a scattering medium. Where the previous state-of-the-art method reports hours to \rev{compute} 
the transmission matrix \rev{on a GPU}, our method takes only a few minutes \rev{on a CPU}.
\end{abstract}
\section{Introduction} \label{sec:intro}

Imaging through a complex optical medium is conceptually simple. The relationship between the lightfield in the input plane $\vx$ and the scattered lightfield $\vy$ is given as
\begin{align}
    \vy = \mA \vx , \label{eq:scattering}
\end{align}
with $\mA$ being the transmission matrix (TM) of the medium (the optical system). Hence, to find the unknown image $\vx$ from measurements $\vy$ it suffices to 
\rev{solve the linear inverse problem \eqref{eq:scattering}}.
Alas, once we set out to do so we quickly realize that 1) the transmission matrix $\mA$ is typically unknown, especially with random scattering media, and 2) $\mA \in \C^{M \times N}$ and $\vy \in \C^M$ in \eqref{eq:scattering} are complex, but typical camera sensors only measure the intensity $|\vy|^2$. Finding a solution for these two difficulties is key to multiple applications such as imaging through fog, paint and tissues in the human body, as well as optics-accelerated signal processing \cite{gupta2019don}.

There are several ways forward. One is to exploit the statistical properties of $\mA$. This is useful in media like tissues or fog, where $\mA$ exhibits certain Gaussian statistics and it enables the design of correlation-based imaging algorithm\rev{s} \cite{hsu2017correlation}. The other way, which gives better images and enables a score of other applications, is to somehow learn $\mA$. Unfortunately, identifying $\mA$ by probing signals in multiply-scattering media is computationally demanding. Since the phase of the measurements is unknown, it amounts to solving a system of quadratic equations. 
Due to this difficulty, state-of-the-art methods report long calibration times even for small $\mA$. As an example, Rajaei \textit{et al.} proposed an approximate message passing algorithm known as prSAMP \cite{rajaei2016robust} which would take days to reconstruct a TM of size $256^2 \times 64^2$ (that is, for input images of size $64 \times 64$). Sharma \textit{et al.} \cite{sharma2019inverse} improve this by bringing the calibration time down to under five hours for a matrix of the same size. 

In this paper we propose a fast numerical interferometry method to rapidly identify $\mA$. Running on our system, our proposed method takes 6 minutes for a calibration problem where the method of Sharma \textit{et al.} \cite{sharma2019inverse} takes 200 minutes (3.29 hours).\footnote{\rev{GPU accelerated} code for Sharma \textit{et al.} downloaded from link in their paper. All parameters were kept the same \rev{and the same GPU was used}.}

The gist of our method is in the special design of the calibration inputs, which allows us to \rev{numerically} find the phases of the measurements $\vy$ by ``multilateration'' in the complex plane, and then \rev{numerically} find $\mA$ by inverting a linear system. Further, instead of direct inversion, we design calibration inputs so that the linear system is solved efficiently by the fast Fourier transform (FFT).

The outline of the algorithm is
\begin{enumerate}
    \item Recover the phase of the calibration measurements by solving a distance geometry problem \cite{gupta2019don};
    \item With the measurement phase recovered, estimate the transmission matrix $\mA$ by solving a linear system;
    \item With $\mA$ in hand, use a phase retrieval method such as Wirtinger flow \cite{candes2015phase} to find $\vx$ from $\abs{\vy}^2$.
\end{enumerate}

We build upon our previous work of casting measurement phase retrieval as a distance geometry problem \cite{gupta2019don}. However, since here we are after speed, we multilaterate all the entries of $\mA$ by solving a suitable linear system.
We verify that the transmission matrices we compute are correct by using them to reconstruct a known input signal. Experiments on real \rev{optical} hardware show that our method works even with quantization and other inevitable hardware noise, and that it only takes a few minutes compared to hours with the previous state of the art.

\subsection{Related work}

In a coherent interferometric setup the phase of $\vy$ can be measured directly, thus immediately giving a linear system to find $\mA$ \cite{popoff2010measuring, yoon2015measuring, popoff2010image, choi2011overcoming}. The downside is that such devices are harder and more expensive to build, often achieve lower frame rates, and are sensitive to environmental factors \cite{dremeau2015reference, yoon2015measuring}.

A numerical alternative with intensity-only measurements is the \textit{double phase retrieval} technique \cite{dremeau2015reference,rajaei2016intensity}. In double phase retrieval the TM is directly estimated from measurements obtained by known probe signals at the input. Calibration is posed as a quadratic phase retrieval problem, with the rows of the TM being the signals to be recovered. \rev{Here} the probe signals \rev{act} as the \rev{observation} 
matrix, \rev{and the TM is the unknown}. Solving these phase retrieval problems can be time consuming even with GPU-accelerated implementations \cite{sharma2019inverse}.

We previously introduced a distance geometry approach to find the phase of $\vy$ \cite{gupta2019don}. In that work the focus was on computing the phase of Gaussian random projections for machine learning tasks such as dimensionality reduction and kernel methods, \rev{without determining $\mA$}. Thus, while directly adapting the approach presented there for calibration is already faster than the double phase retrieval method, here we propose methods that are much faster than both.

Knowing $\mA$ enables \rev{reconstructing} 
$\vx$ from the magnitude of complex measurements $|\vy|^2 = |\mA \vx|^2$ via classical phase retrieval techniques \cite{fienup1982phase, jaganathan2015phase, shechtman2015phase, candes2015phase}. \rev{Furthermore, in} optically-accelerated computing there are uses for fast multiplication of large signals with known random Gaussian iid matrices. Some applications are classification with random features \cite{rahimi2008random}, matrix sketching \cite{tropp2017practical} and randomized linear algebra \cite{halko2011finding}.

\section{Rapid transmission matrix computation}

In this section we describe the procedure outlined in Section \ref{sec:intro} in detail. We adopt the classical system identification paradigm and probe the optical system with a collection of $K$ known calibration signals arranged in the matrix $\mXi = [\vxi_1, \ldots, \vxi_K] \in \R^{N \times K}$,

To convert phase retrieval into multilateration we will also use $S$ known signals $\vv_1, \ldots, \vv_S$, which we ascribe to the columns of $\mV \in \R^{N \times S}$. Without loss of generality we fix $\vv_S$ to be the origin. We will call $\mA \vv_s$ \textit{anchors}.

We now feed the calibration signals through the optical system, forming 
\[
    \mY = \mA \mXi.
\]
The $k$th column of $\mY$ will be denoted $\vy_k = \mA \vxi_k$. We do the same with the differences between \rev{the} columns \rev{of} $\mXi$ and the vectors $\vv_s$, \[
    \vr_s := \mA \vv_s \quad \text{and} \quad \vy_{ks} := \mA(\vxi_k - \vv_s),
\]
for all $k \in \set{1, \ldots, K}$ and $s \in \set{1, \ldots S}$. The $m$th entry of these vectors will be denoted $y_{k,m}$, $r_{s,m}$ and $y_{ks, m}$; the $m$th row of $\mA$ will be denoted $\va^m$.
\rev{Optically, we} can only measure the entrywise magnitudes of these vectors and matrices. The term \textit{numerical interferometry} comes from the fact that the \rev{optically} measured magnitude of $\vy_{ks}$ is the interference pattern between $\mA \vxi_k$ and $\mA \vv_s$.

\subsection{Measurement phase retrieval}

\begin{figure}[t]
\centering
\def\svgwidth{7.25cm}
\fontsize{8}{8}\selectfont
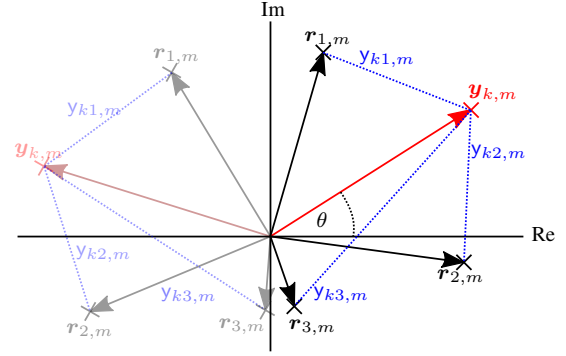
\caption{Two sets of points on the complex plane which are related by \rev{reflection and rotation} transformations. $\vy_{k,m}$ is the \rev{position of} calibration signal $k$ on the complex plane. We can \rev{optically} measure the squared distance between this point and anchor points on the complex plane to obtain the \rev{calibration} signal measurement phase.}
\label{fig:explainer_plane}
\end{figure} 

We consider recovering the phase of the calibration measurements $\mY$ from the system
\begin{align}
    \abs{\mY}^2 = \abs{\mA \mXi}^2. \label{eq:mpr}
\end{align}
We begin by noting that the absolute phase of $\mA \mXi$ in \eqref{eq:mpr} cannot be recovered since conjugating a subset of rows of $\mA$ or adding arbitrary constant phases leads to the same magnitude measurements. This ambiguity is standard in phase retrieval \cite{candes2015phase}.

We remark that for many multiple scattering media, and in particular for the setup used in our experiments, $\mA$ is approximately iid standard complex Gaussian. As such, adding constant phase to rows or conjugating them does not change their distribution. As long as the relative phase between the \textit{columns} of $\mY$ does not change, our method recovers an $\mA$ which is related to the true one by row phasing and conjugation, which is innocuous for many applications.

Without loss of generality we explain the phase retrieval algorithm for the $m$th row of $\mA$. We begin with the trivial observation that the magnitude of a complex number $\mathsf{r}_{s, m} := |r_{s, m}|$ is its distance to the origin. Similarly, the magnitude of a difference between $y_{k, m}$ and $r_{s, m}$, $\mathsf{y}_{ks, m} := |y_{k, m} - r_{s, m}| = |\inprod{\va^m, \vxi_k - \vv_s}|$, is the distance between these numbers in the complex plane (see Fig.~\ref{fig:explainer_plane}).

Suppose for a moment that the points $r_{s, m}$, $s \in \set{1, \ldots, S}$ are known and fixed and that we wish to recover the phase of $y_{k,m}$. By 
\rev{having access to} the distances from $y_{k,m}$ to at least three anchors $r_{s,m}$, we can ``localize'' $y_{k,m}$ by trilateration and hence recover its phase. In practice, having more anchors leads to more robust results.

\rev{To} get a fast method, we now show how the phase of all the entries in the $m$th row of $\mY$ can be computed at once by solving a small linear system with multiple right-hand sides. We adapt a procedure from \cite{stoica2006lecture}. First we expand the squared distance to anchors,
\[
    \sfy_{ks, m}^2 = \sfr_{s, m}^2 + \sfy_{k, m}^2 - 2 r_{s, m}^T y_{k, m},
\]
where we abused notation by interpreting complex numbers as vectors in $\R^2$ that can be transposed. Rearranging, we get
\begin{equation}
    \label{eq:multi_single}
    \sfy_{ks, m}^2 - \sfr_{s, m}^2 = \underbrace{[-2 r_{s, m}^T, \ 1]}_{=: \vm_{s,m}^T} \underbrace{\begin{bmatrix} y_{k,m} \\ \sfy_{k,m}^2 \end{bmatrix}}_{=: \vw_{k,m} \in \R^3}.
\end{equation}
In \eqref{eq:multi_single}, the left-hand side contains only known quantities (anchors are assumed known and $\sfy_{ks,m}^2$ is \rev{optically} measured). The row vector on the right-hand side is also known, while the column vector contains the desired, unknown complex point. 

With $S$ anchors we get $S$ equations for the column vector in \eqref{eq:multi_single}; the system 
\rev{is invertible if}
$S \geq 3$ and the anchors are not colinear. Let \[ \ve_{k, m} = [\sfy_{k1,m}^2 - \sfr_{1,m}^2, \ldots, \sfy_{kS,m}^2 - \sfr_{S,m}^2]^T \] denote the stack of the left-hand sides of \eqref{eq:multi_single} for all $S$ anchors, and $\mE_m = [\ve_{1, m}, \ldots, \ve_{K, m}] \in \R^{S \times K}$ a horizontal stack of $\ve_{k,m}$ for all $K$ calibration signals. Similarly, let $\mM_m = [\vm_{1,m}, \ldots, \vm_{S,m}]^T \in \R^{S \times 3}$ and $\mW_m = [\vw_{1,m}, \ldots, \vw_{K,m}] \in \R^{3 \times K}$. We can then write \eqref{eq:multi_single} for all anchors and calibration signals at once as
\begin{equation}
    \label{eq:multilateration_all}
    \mE_m = \mM_m \mW_m.
\end{equation}
By solving the multiple multilateration problems \eqref{eq:multilateration_all} to get $\wh{\mW}_m = \mM_m^\dag \mE_m$, the top two rows of $\wh{\mW}_m$ contain the real and imaginary parts of all the entries in the $m$th row of $\mY$, as shown in Fig.~\ref{fig:explainer_plane}.

\subsection{Initial anchor positioning}

For the developments of the previous section to make sense, we need to know the anchor positions. Here we propose to follow \cite{gupta2019don} and use classical multidimensional scaling (MDS) summarized below; for a tutorial explanation of MDS see  \cite{dokmanic2015euclidean}.

\rev{We begin} by \rev{optically} measuring the anchors and all their pairwise differences. For row $m$ of $\mA$ for all $(q,s)$ we can get $S(S-1)/2$ squared Euclidean distances between points $\{r_{s,m}\}_{s=1}^S$ on the complex plane,
\begin{align}
    \abs{\inprod{\va^m, \vv_q - \vv_s}}^2 = \abs{r_{q,m} - r_{s,m}}^2 .
\end{align}
We \rev{arrange} these distances into a squared Euclidean distance matrix, $\mD_m \in \R^{S \times S}$, where $d_{qs,m} = \abs{r_{q,m} - r_{s,m}}^2$. Next, denoting the geometric centering matrix $\mJ := \mI_S - \frac{1}{S}\vec{1}_S \vec{1}_S^\T$ where $\vec{1}_S$ is a column vector of $S$ ones, we compute the eigendecomposition of 
\begin{align}
    \mG_m = -\frac{1}{2} \mJ \mD_m \mJ \label{eq:gram}
\end{align}
as $\mG_m = \mU \diag(\lambda_1, \ldots, \lambda_S) \mU^\T$, where the eigenvalue sequence $(\lambda_s)_{s=1}^S$ is nonincreasing. Anchor $s$ is then located on the complex plane at the $s$th element of $(\sqrt{\lambda_1}\vu_1 + j\sqrt{\lambda_2}\vu_2)$ where $\vu_1$ and $\vu_2$ are the first and second columns of $\mU$. As $\vv_S$ was set to be the origin, we subtract the $S$th element of $(\sqrt{\lambda_1}\vu_1 + j \sqrt{\lambda_2}\vu_2)$ from all localized anchors. 

With the anchors localized, we can solve \eqref{eq:multilateration_all} to retrieve the measurement phase for each row. We note that for each row we can apply rotation and reflection transformations to the set of anchor points in the two-dimensional complex plane while still maintaining the \rev{optically} measured distances. As mentioned earlier, rotation and reflection correspond to adding constant phase or conjugating rows.

We note that this approach of measurement phase retrieval varies from the one presented in our earlier work \cite{gupta2019don}. In our earlier work, rather than using \eqref{eq:multilateration_all} to obtain the phase of the measurements, we used MDS to solve a separate distance geometry problem for each element of $\mY$ which is much more time consuming.

\vspace{-3mm}

\section{Obtaining the transmission matrix}
\def \wmA {\mat{\widehat{A}}}

With the phases of $\mY$ computed and recalling that the probe signals $\mXi$ are known, we can \rev{numerically} compute the transmission matrix by solving
\begin{align}
\mY = \mA \mXi  \label{eq:probe_signals}.
\end{align}
The least-squares fit $\wmA = \argmin_{\mA} ~ \norm{\mY - \mA \mXi}_F^2$ is formally given by the pseudoinverse. We design $\mXi$ to ensure that it has full row rank and that the the number of probe signals, $K$, is larger than $N$; then $\wmA = \mY \mXi^\dagger$, where ${}^\dagger$ denotes the pseudoinverse.

Although this method (see Section \ref{sec:experiments}) already gives significant speed gains over the state-of-the-art, for large signals and large $K$, calculating the pseudoinverse can be slow. We next show how to further speed up the algorithm using fast Fourier transforms (FFTs) and a generalized right inverse of $\mXi$ instead of a pseudoinverse.

\subsection{Circulant probes for FFT}

We first note that due to noise and other adversarial effects, it is favorable to have more calibration signals than the minimum number $N$. For simplicity, in this section we choose $K = 2N$. If $\mXi^\dag$ was a circulant matrix, the multiplication $\mY \mXi^\dag$ would be efficiently computed by an FFT as circulant matrices are diagonalized with the discrete Fourier transform (DFT) matrix. With this in mind, we design $\mXi$ instead of $\mXi^\dag$. We use the following proposition\footnote{\rev{For reasons of space we omit the straightforward proof.}} with $\mF$ denoting the unitary DFT matrix and $^*$ denoting the Hermitian transpose:

\begin{proposition}
Let $\mC_1$, $\mC_2$ be invertible circulant matrices of same size, diagonalized as $\mF^*\mLambda_1\mF$ and $\mF^*\mLambda_2\mF$. Then for any $\alpha, \beta$ such that $\alpha + \beta = 1$, $[\alpha\mC_1^{-1}, \ \beta\mC_2^{-1}]^\T$ is a generalized right inverse of $[\mC_1, \ \mC_2]$, with $\mC_1^{-1}$ and $\mC_2^{-1}$ being the circulant matrices $\mF^*\mLambda_1^{-1}\mF$ and  $\mF^*\mLambda_2^{-1}\mF$.
\end{proposition}

We design $\mXi$ as a concatenation of two circulant matrices of size $N \times N$, $\mXi = [\mXi_A, \ \mXi_B] \in \R^{N \times 2N}$. They are diagonalized by the DFT matrix $\mF \in \C^{N \times N}$,
\[
    \mXi_A =  \mF^{*}\mLambda_A \mF \quad \text{and} \quad \mXi_B = \mF^{*}\mLambda_B \mF,
\]
where $\mLambda_A$ and $\mLambda_B$ are diagonal eigenvalue matrices whose entries are the DFT of the first columns of $\mXi_A$ and $\mXi_B$, and $f_{mn} = e^{-j 2\pi mn/N}/\sqrt{N}$. With this notation we can write
\begin{align*}
    \mY &= \mA \begin{bmatrix} \mF^{*}\mLambda_A \mF &  \mF^{*}\mLambda_B \mF \end{bmatrix}. 
\end{align*}
%
%
From here, splitting $\mY$ into halves as $\mY = [\mY_A, \ \mY_B]$ and applying the proposition with $\alpha = \beta = \tfrac{1}{2}$ to obtain a right inverse gives
\begin{align}
    \wmA
    = \frac{1}{2} \left( \mY_A \mF^* \mLambda_A^{-1}  + \mY_B \mF^* \mLambda_B^{-1} \right) \mF \label{eq:fft_method}
\end{align}
Multiplications by $\mF$ and $\mF^*$ can be implemented efficiently using the FFT in time $\calO(N \log N)$. Since $\mLambda_A$ and $\mLambda_B$ are diagonal, multiplying by them takes time $\calO(N)$ and their entries are calculated efficiently using the FFT.




\subsection{Algorithm complexity}

To evaluate our algorithm's complexity, we first consider measurement phase retrieval which is the same process repeated $M$ times. Anchor localization via MDS entails creating $\mG_m$ \eqref{eq:gram} which is $\calO(S^2)$ thanks to the special structure of $\mJ$, and performing a singular value decomposition on it which is $\calO(S^3)$.  Next, with $K = 2N$, solving \eqref{eq:multilateration_all} requires some $\calO(SN)$ operations to obtain $\mE_m$ plus computing a pseudoinverse of an $S \times 3$ matrix which is $\calO(S)$ and finally multiplying the pseudoinverse which is $\calO(SN)$. As these steps are repeated $M$ times, the complexity of measurement phase retrieval is $\calO(MS^3) + \calO(MSN)$. The number of anchors, $S$, is a fixed parameter which gives $\calO(MN)$.

Recovering $\mA$ from $\mY = \mA\mXi$ via \eqref{eq:fft_method} involves three FFT uses with each time on $M$ signals of length $N$ which gives complexity $\calO(MN\log N)$. Multiplications with the diagonal matrices are $\calO(MN)$ and so computing \eqref{eq:fft_method} is $\calO(MN \log N)$. Putting this together with $\calO(MN)$ from measurement phase retrieval results in our algorithm ultimately being $\calO(MN \log N)$. Without using the FFT, it would have been $\calO(N^3) + \calO(MN^2)$.


\section{Experimental verification}
\label{sec:experiments}

We \rev{numerically} compute TMs from real \rev{optical} hardware measurements and use them for \rev{optical} imaging. We use the scikit-learn interface to a publicly available cloud-based optical processing unit (OPU)\footnote{\url{https://www.lighton.ai/lighton-cloud/}. \\
Code \rev{available at \url{https://github.com/swing-research/numerical_interferometry}}.}. The OPU ``imprints'' a signal onto a coherent light beam using a digital micro-mirror device (DMD) which is then shined through a multiple scattering medium such as a white polymer layer. The scattering medium acts like an iid standard complex Gaussian matrix. A camera with 8-bit precision measures the intensity of the scattered light.

\rev{One challenge of using this setup is that the DMD only produces binary inputs}. \rev{Therefore,} the anchor signals have to be designed so that the difference between any two anchors is binary \rev{and they} can be used to localize anchors \rev{on the complex plane}. Furthermore, the probe signals, $\mXi$, should be binary and remain binary when any anchor signal is subtracted from them. We design anchors by summing all probe signals and existing anchors, thresholding indices with values greater than one to one and then making 15\% of the values at indices where the sum was zero to one.

From \eqref{eq:probe_signals} we need all rows of $\mXi$ to have at least one nonzero value in order to reconstruct all columns of $\mA$. However, this results in anchors which are all one. We therefore apply our method twice with two different \rev{sets of probe signals} with different zero rows. Each set of probe signals recovers a subset of the columns of the TM. \rev{Since the two sets of probe signals share some non-zero row indices, the corresponding recovered columns should be identical. We use this fact to rotate and conjugate the rows of the two recovered TMs so that the common columns coincide.} Finally columns of $\mA$ that are only recovered by one \rev{set of probing signals} are collected to form a complete TM.

\rev{More concretely, each $\mXi \in \R^{N \times K}$ is built by concatenating two circulant matrices, each of size $K/2 \times K/2$. The entries of the non-zero columns of circulant matrices are drawn iid from Bernoulli$\left(\tfrac{1}{2}\right)$ and $N-(K/2)$ all-zero rows are inserted at random indices so that $\mXi$ is $N \times K$. When inverting $\mXi$ to calibrate the TM, we remove the inserted zero rows to get block circulant matrices and use \eqref{eq:fft_method} to recover a subset of the columns of $\mA$.}

\subsection{\rev{Optical} imaging with \rev{numerically computed} TMs}

\begin{table}[t]
\centering
\begin{tabular}{@{}cccc@{}}
\toprule
$\boldsymbol{N}$ & $\boldsymbol{M/N}$ & \textbf{\begin{tabular}[c]{@{}c@{}}Time taken\\ (minutes)\end{tabular}} & \textbf{\begin{tabular}[c]{@{}c@{}}Reconstructed image\\ relative error\end{tabular}} \\ \midrule
$32^2$ & 32 & 0.97 & 10.2\% \\
$32^2$ & 64 & 2.05 & 7.8\% \\
$32^2$ & 128 & 4.01 & 6.2\% \\ \midrule
$64^2$ & 16 & 6.15 & 21.5\% \\
$64^2$ & 32 & 11.69 & 15.3\% \\
$64^2$ & 64 & 24.14 & 12.0\% \\ \midrule
$96^2$ & 16 & 31.36 & 25.8\% \\ \midrule
$128^2$ & 12 & 71.97 & 32.0\% \\ \bottomrule
\end{tabular}
\caption{The time taken to reconstruct TMs, $\mA \in \C^{M \times N}$ and the relative error of reconstructed images using these TMs. All times were measured on the same system.}
\label{table:times_errors}
\vspace{-0.95em}
\end{table}

We reconstruct images of different sizes and vary the oversampling factor, $M/N$, of the \rev{optically-obtained} quadratic \rev{magnitude} measurements which results in different TM sizes. \rev{Wirtinger flow with random uniform initialization is used for reconstruction \cite{candes2015phase, chen2018gradient}. 
During gradient descent, any} elements of the signal with absolute value greater than one are normalized.
500 iterations of gradient descent are done for $32 \times 32$ images. For the other images, the absolute value of the recovered signal after 500 initial iterations with $4N$ measurements is used to start 2000 further iterations which includes all measurements.
20 anchor signals are used for all images except the $128 \times 128$ ones which use 15. For $32 \times 32$ and $64 \times 64$ images $K = 1.5N$, for $96 \times 96$, $K = 1.125N$ and for $128 \times 128$, $K = 1.03125N$.

\begin{figure}[t]
    \centering
    \vspace{-0.9em}
    \includegraphics[width=0.8\linewidth ]{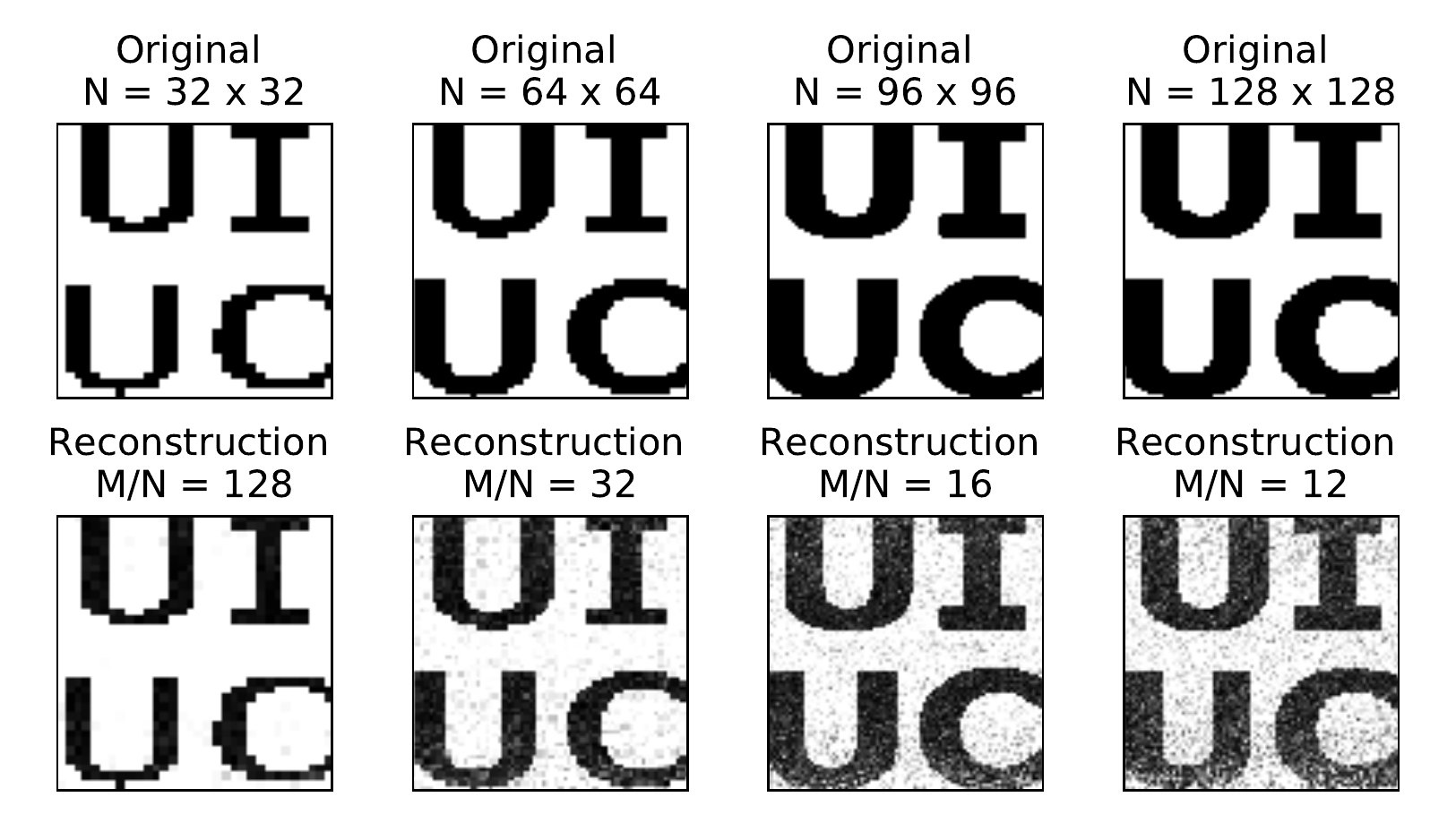}
    \vspace{-1.25em}
    \caption{Reconstructions with some of the transmission matrices reported in Table \ref{table:times_errors}. The top row shows the original and the bottom row shows the absolute value of its corresponding reconstruction.}
    \label{fig:recons}
\end{figure}

Table \ref{table:times_errors} shows the TM calibration time as well as the reconstructed image relative error which we define as \rev{$\|{\hat{\vx} - \vx}\|_2 / \norm{\vx}_2$} where $\hat{\vx}$ is the reconstructed signal. \rev{We note that we are doing phase retrieval with noise as the TM has been estimated and an 8-bit precision camera obtains optical measurements.} The error decreases with increasing oversampling factor which matches the theory. As mentioned in Section \ref{sec:intro}, a $256^2 \times 64^2$ TM took 3.26 hours when prVAMP was used \cite{sharma2019inverse}. In contrast our method takes 6.15 minutes. In fact the biggest TM listed in Table \ref{table:times_errors} is 12 times the size of this TM. Fig. \ref{fig:recons} displays some reconstructions using our calibrated TMs.

\subsection{Computation time scaling}

\begin{figure}[t]
    \centering
    \includegraphics[width=.8\linewidth]{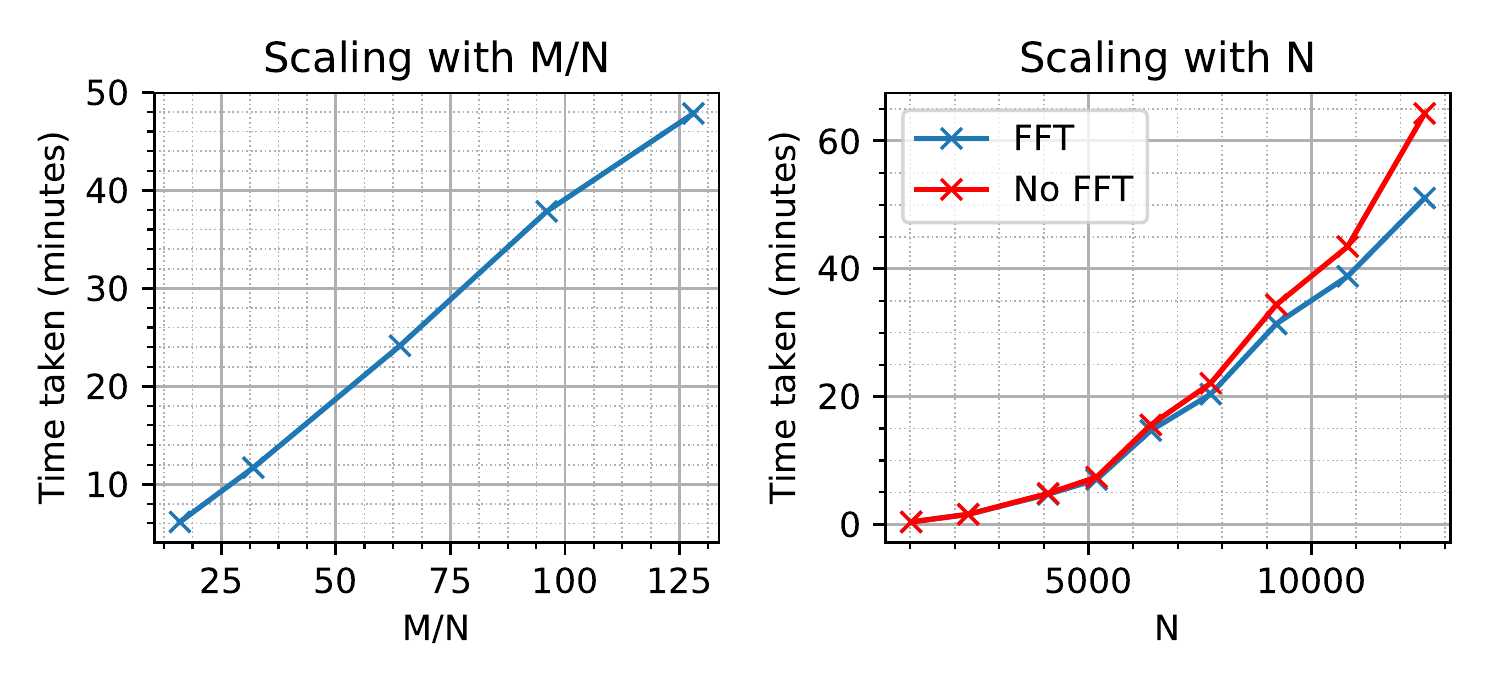}
    \vspace{-1.25em}
    \caption{(Left) Time taken to calibrate TM when the oversampling factor is varied; (Right) Time taken when input dimension varies and oversampling factor is fixed at 16.}
    \vspace{-0.95em}
    \label{fig:scaling}
\end{figure}

From Table \ref{table:times_errors} and Fig. \ref{fig:scaling} (left) we can see that the time taken increases linearly with the oversampling factor. In Figure \ref{fig:scaling} (left) $N = 64^2$, $K = 1.5N$ and 20 anchor signals are used. In Figure \ref{fig:scaling} (right), the oversampling factor is fixed at 16, $K = 1.125N$, 20 anchors are used and the input dimension is varied. As signal dimension increases, using the FFT  to solve \eqref{eq:probe_signals} yields increasing gains.

\section{Conclusion}

Calibrating a transmission matrix with intensity-only measurements has traditionally been a time consuming process. In this work we propose a method to reduce the calibration time for typical size matrices from hours to minutes. The crux of our method is the rapid recovery of the measurement phase due to probe input signals. With the obtained measurement phase and probe signals, transmission matrix calibration then amounts to solving a simple linear system which we do efficiently with the FFT. Experiments on optical hardware confirm that our method is indeed faster than existing methods and reconstructs transmission matrices which can be used for imaging.


\end{document}